\pdfoutput=1

\documentclass[11pt]{article}

\usepackage[final]{acl}

\usepackage{times}
\usepackage{latexsym}
\usepackage{amsmath}
\usepackage{float}
\usepackage{booktabs}
\usepackage{multirow}

\usepackage{subcaption}
\usepackage{cuted}
\usepackage{soul}  
\usepackage{xcolor} 

\usepackage{pdfpages}
\usepackage{algorithm}
\usepackage{algpseudocode}
\usepackage[most]{tcolorbox}

\usepackage{hyperref}
\usepackage{nameref}

\usepackage{afterpage}

\usepackage{booktabs}

\definecolor{hlblue}{rgb}{0.4,0.6,0.8}
\definecolor{hlred}{rgb}{0.8,0.2,0.2}
\definecolor{hlpurple}{rgb}{0.5,0.2,0.5}
\definecolor{hlorange}{rgb}{1.0,0.5,0.0}
\definecolor{hlteal}{rgb}{0.0,0.5,0.5}
\definecolor{darkgreen}{rgb}{0.0, 0.5, 0.0}

\newcommand{\hlblue}[1]{\textcolor{hlblue}{#1}}
\newcommand{\hlred}[1]{\textcolor{hlred}{#1}}
\newcommand{\hlpurple}[1]{\textcolor{hlpurple}{#1}}
\newcommand{\hlorange}[1]{\textcolor{hlorange}{#1}}
\newcommand{\hlteal}[1]{\textcolor{hlteal}{#1}}
\usepackage[T1]{fontenc}

\usepackage[utf8]{inputenc}

\usepackage{microtype}

\usepackage{inconsolata}

\usepackage{graphicx}

\usepackage{listings}
\usepackage{todonotes} 

\definecolor{BLUE}{RGB}{30,60,220} 
\definecolor{GREEN}{RGB}{0,0,64} 
\definecolor{titleblue}{RGB}{47,84,150}

\title{
\textsc{\textbf{
\textcolor{titleblue}{\textsc{Technique}}\textcolor{titleblue}{\textsc{RAG}}}}: Retrieval Augmented Generation for Adversarial Technique Annotation in Cyber Threat Intelligence Text}

\author{
  \textbf{Ahmed Lekssays\textsuperscript{1}},
  \textbf{Utsav Shukla\textsuperscript{2}},
  \textbf{Husrev Taha Sencar\textsuperscript{1}},
  \textbf{Md Rizwan Parvez\textsuperscript{1}}
\\
\\
  \textsuperscript{1}Qatar Computing Research Institute, Doha, Qatar,
  \textsuperscript{2}Independent Researcher
\\
  \small{
    \texttt{alekssays@hbku.edu.qa, utsavshuk@gmail.com, hsencar@hbku.edu.qa, mparvez@hbku.edu.qa}
  }
}

\usepackage{multirow}

\usepackage{pgfplots}
\pgfplotsset{compat=1.17}

\definecolor{chart_specific_models}{RGB}{217, 240, 211}

\newcommand{\model}{\textsc{TechniqueRAG}~}
\newcommand{\modelnospace}{\textsc{TechniqueRAG}}

\usepackage{subcaption}
\usepackage{booktabs}
\usepackage{multirow}
\usepackage{tabularx}
\usepackage{longtable}

\usepackage{pgfplots}
\pgfplotsset{compat=1.17}
\usepackage{pgfplotstable}
\usepackage{tikz}

\usepackage{graphicx}
\graphicspath{ {./images/} }

\begin{document}
\maketitle
\begin{abstract}

Accurately identifying adversarial techniques in security texts is critical for effective cyber defense. However, existing methods face a fundamental trade-off: they either rely on generic models with limited domain precision or require resource-intensive pipelines that depend on large labeled datasets and task-specific optimizations—such as custom hard-negative mining and denoising—resources rarely available in specialized domains. We propose \textbf{\modelnospace}, a domain-specific retrieval-augmented generation (RAG) framework that bridges this gap by integrating off-the-shelf retrievers, instruction-tuned LLMs, and minimal text–technique pairs. First, our approach mitigates data scarcity by fine-tuning only the generation component on limited in-domain examples, circumventing resource-intensive retrieval training. Second, although conventional RAG mitigates hallucination by coupling retrieval and generation, its dependence on generic retrievers often introduces noisy candidates, thereby limiting domain-specific precision. To address, we  enhance the retrieval quality and domain specificity through a zero-shot LLM re-ranking that explicitly aligns retrieved candidates with adversarial techniques. 
Experiments on multiple security benchmarks demonstrate that \model achieves state-of-the-art performances without extensive task-specific optimizations or labeled data, while comprehensive analysis provides further insights.
\end{abstract}

\section{Introduction}

\begin{figure}[t]
\begin{tcolorbox}[title=Example of text to {\em (sub-)techniques} annotated pairs, colback=blue!5!white, colframe=blue!25!black, fonttitle=\footnotesize, boxrule=0.2mm, sharp corners]
\small
  Monero miner scripts are \hlblue{downloaded from TeamTNT's server} and piped to \hlred{``bash''} using a \hlpurple{SSH session} as the \hlorange{``root'' user} with \hlteal{private key from ``/tmp/TeamTNT.''} \\
---------------------------------------------------------------  \\
1. \hlorange{T1098.004: SSH Authorized Keys}\\
2. \hlblue{T1195: Supply Chain Compromise}\\
3.  \hlred{T1059.004: Unix Shell}\\
4.  \hlpurple{T1021.004: Remote Services: SSH}\\
5.  \hlteal{T1552.004: Private Keys}
\end{tcolorbox}
\vspace{-12pt}
\caption{Example of MITRE ATT\&CK {\em techniques} and {\em sub-techniques} highlighted in text with corresponding colored (implicit) indicators. IDs with "." denote sub-{\em techniques} (e.g., T1098.004).}
\label{fig:mitre_example}
\vspace{-20pt}
\end{figure}

Uncovering new adversarial behaviors is critical for strengthening defenses against rapidly evolving cyber threats. These behaviors, defined by the tools, techniques, and procedures used by attackers, reveal how adversaries plan and execute attacks and impact systems and data. By identifying and analyzing the traces or artifacts left behind, security analysts can map low-level actions to higher-level concepts, such as \emph{tactics} (i.e., strategic objectives like "lateral movement"), \emph{techniques} or fine-grained \emph{sub-techniques} (i.e., tactical methods like "Debugger Evasion"), and \emph{procedures} (i.e., implementation details like "using PowerShell for credential dumping"). 
The findings are shared with other experts through public channels and threat intelligence services via security reports and detailed descriptions, to strengthen defenses, anticipate possible threats, and improve incident response.

The MITRE ATT\&CK framework has established itself as the industry standard for categorizing and mapping adversarial behaviors \cite{mitre-attack}. 
This framework provides a comprehensive knowledge base of adversarial \textit{tactics, techniques}, and \textit{procedures (TTPs)}, built from real-world threat intelligence and incident response data. It uses a hierarchical matrix structure to systematically organize and classify adversary behaviors. 
The broad adoption of the ATT\&CK framework presents a significant operational challenge: security analysts must manually map ambiguous threat descriptions from incident reports (such as the example shown in Fig \ref{fig:mitre_example}) to standardized ATT\&CK \emph{(sub)-techniques}—a time-intensive process that demands expert knowledge.
This manual task has motivated research into automated adversarial \emph{technique} identification, which aims to systematically label text segments with their corresponding ATT\&CK \emph{technique} and \emph{sub-technique} IDs.

Prior approaches for \emph{(sub-)technique} annotation adopt two primary paradigms: (1) Multi-class classification  that directly map text to \emph{(sub-)technique} IDs \cite{you2022tim,tram:2023}, which, while straightforward to implement, struggle with class imbalance and require extensive labeled training data; and (2) Retrieval/ranking approaches that evaluate semantic similarity between the text and \emph{(sub-)techniques}. Early methods like Ladder \cite{ladder:alam:2022} \& AttackKG \cite{li2022attackg} introduce basic similarity-based ranking. Text2TTP \cite{kumarasinghe2024semantic} advanced this through hierarchical re-ranking with fine-tuned embeddings, while NCE \cite{nguyen-etal-2024-noise-huawei} improved similarity learning using dual-encoder architectures. Recently, IntelEX \cite{xu2024intelex} employed LLMs in both retrieval and zero-shot learning settings to assess \emph{(sub-)technique} relevance.

Although promising, the current methods are constrained by a critical trade-off: they either rely on general-purpose models lacking domain expertise or require large-scale labeled datasets and computationally intensive training pipelines. Retrieval approaches, for instance, require extensive hard-negative mining to distinguish fine-grained {\em (sub-)techniques} 
while classification models demand curated, balanced training sets—resources rarely available in this specialized domain. Compounding this issue, despite MITRE ATT\&CK framework defines over 550 adversarial \emph{(sub-)techniques}, only approximately 10,000 annotated examples are publicly available \cite{kumarasinghe2024semantic,nguyen-etal-2024-noise-huawei}, severely limiting generalization.

\begin{figure}[!t]
\captionsetup[subfigure]{labelformat=empty}
\centering
\includegraphics[width=1.0\linewidth]{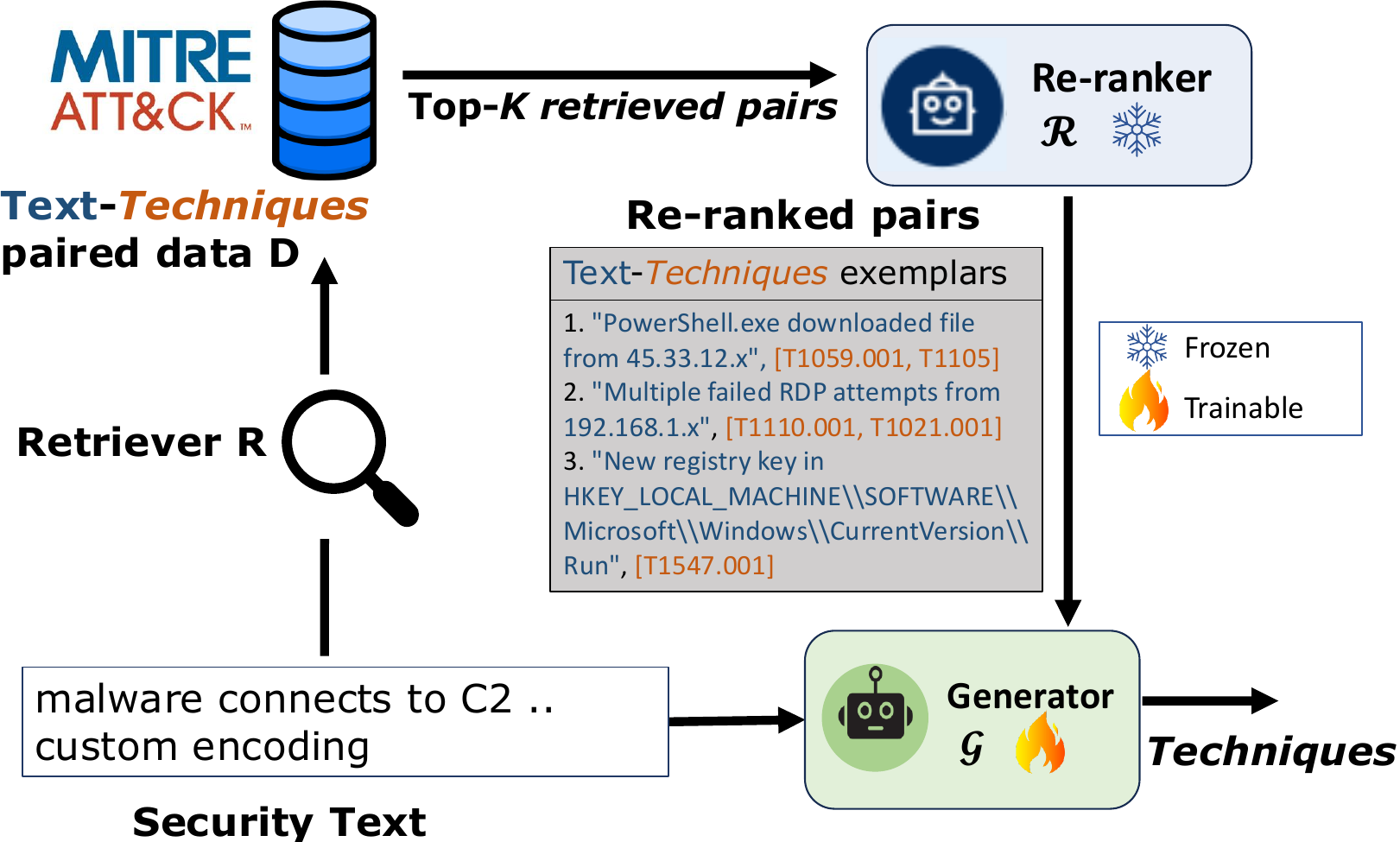}
\vspace{-20pt}
\caption{
Overview of \modelnospace.
}
\vspace{-18pt}
\label{figure:model-overview}
\end{figure}

To address these dual challenges, we propose \textbf{\modelnospace}, a domain-specific retrieval-augmented generation (RAG) framework for \emph{(sub-)techniques} annotation task  that bridges generic and specialized models while eliminating dependency on resource-intensive pipelines or large labeled data. Unlike conventional approaches, \model integrates three key components: (a) off-the-shelf retrievers for candidate extraction (b) instruction-tuned LLMs to re-rank candidate \emph{(sub-)techniques} (c) minimal text-\emph{(sub-)technique} pairs used exclusively for fine-tuning the generator. 
Our approach mitigates data scarcity by fine-tuning only the generation component on limited in-domain examples while leveraging a novel re-ranking framework that uses generic off-the-shelf LLMs without fine-tuning to explicitly align retrieved candidates with adversarial \emph{(sub-)techniques}, thereby enhancing domain specificity. 
Fig \ref{figure:model-overview} shows an overview.

\begin{table*}[t]
\centering
\small
\begin{tabular}{p{3.5cm}p{3.5cm}p{7cm}}
    \toprule
    \textbf{Method} & \textbf{Problem Formulation} & \textbf{Key Features} \\
    \midrule
    TRAM~\cite{tram:2023} & Classification & Utilizes n-gram frequency features \\
    TTPDrill~\cite{husari2017ttpdrill} & Matching/Ranking & Employs TF-IDF and BM25 for text retrieval \\
    AttacKG~\cite{attackg:esorics:2022} & Matching/Ranking & Leverages knowledge graph representations \\
    TIM~\cite{you2022tim} & Classification & Incorporates textual and lexical features \\
    LADDER \cite{ladder:alam:2022} & Matching/Ranking & Uses sentence encoder embeddings \\
    NCE ~\cite{nguyen-etal-2024-noise-huawei} & Matching/Ranking & Applies task-adapted  dual-encoder embeddings \\
    Text2TTP \cite{kumarasinghe2024semantic} & Matching/Ranking & Enhances  dual-encoder retrievals with a cross-encoder embedding filter model \\
    IntexEX \cite{xu2024intelex} & Hybrid (Retrieval and Evaluation) & Combines sentence/entity-based search with LLM-based evaluation of candidate outputs \\
   \midrule
    TechniqueRAG (Ours) & Retrieval Augmented Generation (RAG) & Integrates any retriever with LLM-based re-ranking and fine-tuned generation \\
    \bottomrule
\end{tabular}
\vspace{-5pt}
\caption{Overview of methods proposed for automatically mapping security text to Mitre ATT\&CK \textit{(sub-)technique.}. Our proposed \model leverages a flexible RAG framework by combining off-the-shelf retrievers, a novel LLM-based re-ranking mechanism, and a fine-tuned generator, distinguishing it from prior approaches.}
\label{tab:techniques-methods}
\vspace{-12pt}
\end{table*}

While LLMs offer promising capabilities for ranking adversarial \emph{(sub-)techniques}, this task poses challenges beyond their standard pre-training and alignment objectives. Unlike traditional ranking tasks—such as those encountered in question-answering—security technique ranking requires distinguishing among subtly different ATT\&CK {\em (sub-)technique}  that may co-occur in a text without explicit indicators (see Fig ~\ref{fig:mitre_example}). Consequently, conventional re-ranking frameworks like RankGPT \cite{sun-etal-2023-chatgpt}, though effective for general search, struggle with the nuanced demands of security-specific ranking. To address, our framework prompts LLMs to engage in explicit, step-by-step reasoning about {\em (sub-)technique} relevance, considering both high-level \emph{techniques} and fine-grained \emph{sub-techniques}. This structured decomposition not only enables more precise ranking but also captures hierarchical relationships among \emph{(sub-)techniques} as in the ATT\&CK framework.

In experiments, we evaluate our framework on three benchmarks, addressing both single-label and multi-label prediction tasks for \emph{(sub-)techniques}. Results demonstrate that \model significantly outperforms various baseline approaches, including classification-based, retrieval/ranking-based, and hybrid methods. Furthermore, it achieves comparable or superior performance to vanilla RAG approaches, even when utilizing powerful LLMs like GPT-4o.

\section{Related Work} 
\label{sec:related_work}
Table \ref{tab:techniques-methods} presents a comparison of the methods proposed for the {\em (sub-)technique} ID annotation task. These methods can be categorized into three groups based on their characteristics

\textbf{Text-based Feature Extraction} Initial approaches to {\em (sub-)technique} identification utilized classical text representations: bag-of-words models utilizing TF-IDF \cite{legoy2020automated, tsai2020cti}, n-gram frequencies \cite{legoy2020automated, tram:2023}, and word embeddings \cite{legoy2020automated} as features for multi-class \& multi-label classifiers. Later works enhanced representation through syntactic parsing to extract $(subject, verb, object)$ patterns \cite{ttpdrill:acsac:2017} \& knowledge graph alignment \cite{attackg:esorics:2022} to capture contextual relationships in threat behaviors.

\textbf{Neural Text Embedding Approaches} Transformer-based language models \cite{sbert:corr:2019}
enabled semantic similarity-based technique identification through neural embeddings.  
Early approaches used pre-trained encoders to embed threat behaviors, either handling multi-sentence descriptions \cite{you2022tim} or specific attack patterns \cite{ladder:alam:2022}, evaluating relevance through embedding similarity. 
\cite{kumarasinghe2024semantic} advanced this through a multi-stage architecture combining fine-tuned cross-encoder and dual-encoder models to balance effectiveness and efficiency. \cite{nguyen-etal-2024-noise-huawei} further developed this approach using a dual-encoder architecture with alignment components, leveraging both scratch-trained embeddings and domain-specific models.

\textbf{LLM Applications} The application of LLMs to technique identification has yielded important insights. \cite{kumarasinghe2024semantic} demonstrated that LLMs perform poorly compared to fine-tuned smaller models due to hallucination issues. To address, \cite{xu2024intelex} introduced a hybrid approach combining zero-shot classification, retrieval, and LLM-based validation while we propose RAG
to enhance reliability and reduce errors.

\section{Method}
\label{sec:methodology}

We present \modelnospace, a domain-specific retrieval-augmented generation (RAG) framework for adversarial {\em technique} (or sub-{\em technique}) annotation. Unlike conventional approaches that rely on task-specific optimizations and extensive labeled data, \model effectively integrates retrievers, instruction-tuned LLMs and small-scale text-\emph{techniques} paired data. 
We first provide an overview of our approach, followed by details on the key components: (i) retriever and (ii) LLM-based re-ranking (ii) generator fine-tuning.
Fig \ref{figure:model-overview} presents an overview of our system.


\subsection{Problem Formulation and Overview}
\label{subsec:problem-formulation}




Let \( X = \{ x_1, x_2, \dots, x_n \} \) denote a set of security texts (e.g., attack behaviors) and \( Y = \{ y_1, y_2, \dots, y_m \} \) denote the set of adversarial \emph{(sub-)techniques} (e.g., MITRE ATT\&CK IDs). For a given security text, its annotation is represented as a sequence 
\(
\mathbf{Y} = (y_1, y_2, \dots, y_l),
\)
where each \( y_i \in Y \) and \( l \leq m \). We define \( Y^* \) as the set of all finite sequences over \( Y \), so that \( \mathbf{Y} \in Y^* \).
We assume access to a small paired dataset \( D = \{ (x_1, \mathbf{Y}_1), (x_2, \mathbf{Y}_2), \dots, (x_n, \mathbf{Y}_n) \} \) of threat descriptions and the corresponding set of ground-truth \emph{(sub-)techniques}. 
Given an input text \( x \in X \), the task is to predict the corresponding \emph{(sub-)techniques} \( \mathbf{Y}_x \subseteq Y^* \).


Our framework, \modelnospace, comprises three modules: (1) a \emph{retriever} \( R \), (2) an LLM-based \emph{re-ranker} \( \mathcal{R} \), and (3) a \emph{generator} LLM \( \mathcal{G} \). Given an input \( x^q \), the retriever \( R \) first retrieves the top-\( K \) relevant pairs $R_x$ from the dataset \( D \) based on their similarity to the query text. The re-ranker $\mathcal{R}$ processes the retrieved pair of annotated text, \( R_x\), to produce an ordered sequence $\mathcal{R}_x$, which is then augmented with the input sequence \( x \) to form the generator context
\(
\mathcal{C}_x = x \oplus \mathcal{R}_x
\)
where \( \oplus \) denotes concatenation. To conform to the context length of the generator \( \mathcal{G} \), the user may select \( k \leq K \) re-ranked pairs for augmentation. These augmented pairs serve as exemplars that guide the generation process and help to reduce hallucination. Finally, the generator \( \mathcal{G} \) produces the target output \( \mathbf{Y}_x \) from the augmented input \( \mathcal{C}_x \) (See Fig \ref{figure:model-overview}).
In the following subsections, we provide detailed descriptions of the retriever \( R \), the LLM-based re-ranker \( \mathcal{R} \), and the generator \( \mathcal{G} \) used in \modelnospace.

\subsection{Retriever \(R\)}
\label{sec:retriever}

The retriever module processes a query security text \( x^q \) by leveraging a retrieval corpus \( D_R \) to fetch most relevant candidate pairs. In our setting, due to the lack of additional data, we employ the paired dataset \( D \) both as the retrieval corpus \( D_R \) and as the training data for the generator \(\mathcal{G}\). To prevent data leakage during its training, we explicitly exclude \( x^q \) from \( D_R \), defining it as:
\[
D_R = \left\{ (x_i, \mathbf{Y}_i) \mid (x_i, \mathbf{Y}_i) \in D \land x_i \neq x^q \right\}.
\]
The retriever $R$ returns the top-$K$ pairs 
\( 
 R_x = \{ (x^{R}_{1}, \mathbf{Y}^{R}_{1}), (x^{R}_{2}, \mathbf{Y}^{R}_{2}), \dots, (x^{R}_{K}, \mathbf{Y}^{R}_{K}) \} \subset D_R,
\)
where each pair \((x^{R}_i, \mathbf{Y}^{R}_i)\) corresponds to a security text and its associated {\em (sub-)techniques} from \( D_R \) along with their (lexical/semantic) similarity $sim(x^q, x^{R}_{i}) \geq sim(x^q, x^{R}_{j})\text{ }\forall\text{ }j>i$.
Any off-the-shelf retriever (e.g., sparse: BM25 or dense: pre-trained sentence embedding model) can be employed as $R$.
While our approach is generic and further domain adaptation of $R$ may improve performance, it is important to note that $D$ only has behavior description and \emph{(sub-)technique}  annotation pairings ($x_i, \mathbf{Y}_i$) 
without specifying the absolute relevance of $x_i$ with any of the individual \emph{(sub-)techniques} within the sequence of ground truth technique annotations $\mathbf{Y}_i$. As a result, training $R$ solely with heuristic losses, such as in-batch negatives, leads to sub-optimal adaptation. Furthermore, no hard negatives or denoising data are available. Therefore, instead of fine-tuning a retriever, we employ an off-the-shelf retriever as  $R$  and enhance it through re-ranking, as detailed below.

\subsection{Re-Ranker \(\mathcal{R}\)}
\label{sec:re-ranker}

To address data scarcity and enhance domain-specific precision, our re-ranker \(\mathcal{R}\) refines the candidate set retrieved by \(R\) using an instruction-tuned large language model. Unlike generic prompting frameworks for ranking (e.g., RankGPT~\cite{sun-etal-2023-chatgpt}), which lack domain-specific knowledge, our re-ranker employs a novel prompting framework specifically designed for adversarial technique annotation that addresses three key challenges described below.

\noindent \textbf{Explicit Reasoning for Implicit Mapping:} Security texts rarely provide explicit rationales for technique mappings. For example, the text ``malware connects to C2 using custom encoding'' implies both command-and-control (TA0011) and defense evasion (TA0005) tactics, but doesn't directly state this relationship. \(\mathcal{R}\) instructs the LLM to decompose such implicit connections through structured reasoning:
\begin{tcolorbox}[title=Prompt and Response: Break Down the Query, colback=blue!5!white, colframe=blue!25!black, fonttitle=\footnotesize, boxrule=0.2mm, sharp corners]
\label{fig:prompt_response_break_down_example}
\small
   \# Decompose the given security query into distinct attack steps or phases. \\ 
   \# Identify any implied or explicitly mentioned behaviors that indicate 
     specific adversarial {\em (sub-)techniques}. \\
---------------------------------------------------------------  \\
{Query: ``malware connects to C2 .. custom encoding''}\\
---------------------------------------------------------------  \\
{Step 1: Identify core behavior → C2 communication with encoding}\\
{Step 2: Map to tactics → Command and Control + Defense Evasion}\\
{Step 3: Link to techniques → T1071 (C2 Protocol) + T1027 (Obfuscation)}
\end{tcolorbox}

\noindent \textbf{Balanced Consideration of Multiple Techniques:} Security activities often involve multiple {\em techniques} simultaneously. $\mathcal{R}$ prompt ensures comprehensive coverage through parallel evaluation by instructing to explore each possible {\em technique}. 

\begin{tcolorbox}[title=Prompt and Response: Multiple {\em Techniques}, colback=blue!5!white, colframe=blue!25!black, fonttitle=\footnotesize, boxrule=0.2mm, sharp corners]

\label{fig:prompt_response_multitech_example}
\small
\# Consider that the query may involve multiple  {\em (sub-)techniques}. 
     (both direct and implied). \\
---------------------------------------------------------------  \\
{Query: ``malware connects to C2 .. custom encoding''}\\
---------------------------------------------------------------  \\
{Primary  Technique: T1071 (Application Layer Protocol)}\\
{Rationale: Direct C2 communication behavior}\\
{Secondary  Technique: T1027 (Obfuscated Files or Information)}\\
{Rationale: Custom encoding for evasion}
\end{tcolorbox}


\noindent \textbf{Fine-Grained Relevance w.r.t Sub-Techniques:} Certain \emph{techniques} have multiple sub-\emph{techniques} and some do not. Our framework evaluates subtle distinctions among them to inform the overall ranking of different \emph{techniques} (with and without \emph{sub-techniques}). For instance:

\begin{tcolorbox}[title=Prompt and Response: Fine-grained Relevance, colback=blue!5!white, colframe=blue!25!black, fonttitle=\footnotesize, boxrule=0.2mm, sharp corners]
\small
\# Map each identified step or behavior to the most appropriate technique or sub-technique (if available for each corresponding \emph{technique}). For each matching technique, explain the connection between the query 
     and the corresponding adversarial behavior.\\
---------------------------------------------------------------  \\
{Query: ``PowerShell script encoded in base64 downloads malware from remote server''}\\
---------------------------------------------------------------  \\
{Technique Analysis:}\\
1. T1059 (Command and Scripting Interpreter) \\ 
   - T1059.001 (PowerShell): Direct match for script execution \\ 
   - Confidence: High due to explicit PowerShell usage \\
2.  T1027 (Obfuscated Files) \\ 
   - No sub-techniques apply to basic encoding \\ 
   - Confidence: Med. as common obfuscation method \\
---------------------------------------------------------------  \\
{Final Ranking: T1059.001 > T1027}\\
{Rationale: Sub-technique analysis reveals credential access as primary intent with process injection as supporting mechanism}
\end{tcolorbox}

The complete system prompt to guide the LLM through this structured analysis is provided in Appendix \ref{apdx:prompts}. This hierarchical, reasoning-based approach enables \(\mathcal{R}\) to reorder candidates while maintaining alignment with ATT\&CK's taxonomy, addressing ambiguities in initial retrieval. To address the challenge of processing large candidate sets within the LLM's context window, we  utilize the sliding window mechanism as in \citet{sun-etal-2023-chatgpt}.

\subsection{Generator \(\mathcal{G}\)}
\label{sec:generator}

To adapt the LLM generator \(\mathcal{G}\) that produces the final annotations of the adversarial technique \(\mathbf{Y}_x\) from an augmented input \(\mathcal{C}_x\), we fine-tune it using $D$ as the training set.  
As discussed in Section \ref{subsec:problem-formulation}, the augmentation process concatenates the original query \(x\) with the re-ranked candidate pairs  \(\mathcal{R}_x\) (i.e., $ \ x \ \oplus \mathcal{R}_x $), specifically  as following:
\begin{equation*}
\setlength{\belowdisplayskip}{5pt}
\setlength{\abovedisplayskip}{5pt}
\label{eq:common}
    \begin{split}
        \mathcal{C}_x 
        =& \ x \ [text]  \ x^{\mathcal{R}}_{1} \ [technique] \ \mathbf{Y}^{\mathcal{R}}_{1} \ [text] \\
        & x^{\mathcal{R}}_{2} \ [technique] \ \mathbf{Y}^{\mathcal{R}}_{2} \ldots  \ [text] \\
        &  \ x^{\mathcal{R}}_{k} \ [technique] \ \mathbf{Y}^{\mathcal{R}}_{k},
    \end{split}
\end{equation*}
where ``[]'' denotes separator tokens, $x_{j}$ and $\mathbf{Y}_{j}$ are parallel data (e.g., 
$x_{j}$ is the security text and $\mathbf{Y}_{j} = (y_{1,j}, y_{2,j}, \dots y_{m,j})$ is the corresponding \emph{(sub-)technique} sequence.

We train the generator model \(\mathcal{G}\) minimizing the negative log-likelihood of the ground-truth  \emph{(sub-)technique} annotations  \(\mathbf{Y}_x = (y_{1,x}, y_{2,x}, \dots, y_{l,x})\) conditioned on the augmented input \(C_x\):
\[
\mathcal{L} = - \sum_{(x, \mathbf{Y}_x) \in D} \sum_{i=1}^{l} \log P_{\mathcal{G}}(y_{i,x} \mid \mathcal{C}_x).
\]
To mitigate hallucination, \(\mathcal{G}\) is constrained to generate outputs from the re-ranked candidate set \(C_x\). This design ensures that the final predicted \emph{(sub-)techniques} are both contextually grounded in \(x\) and consistent with the adversarial taxonomy provided by the exemplars in \(C_x\).

\begin{table*}[t]
\centering
\small
\resizebox{0.85\linewidth}{!}{%
\begin{tabular}{l ccc | ccc | ccc}
\toprule
\multirow{2}{*}{\textbf{Model}} & \multicolumn{3}{c|}{\textbf{Tram (Single-label)}} & \multicolumn{3}{c|}{\textbf{Procedures (Single-label)}} & \multicolumn{3}{c}{\textbf{Expert (Multi-label)}} \\
\cmidrule(lr){2-10}
 & \textbf{Prec.} & \textbf{Rec.} & \textbf{F1} & \textbf{Prec.} & \textbf{Rec.} & \textbf{F1} & \textbf{Prec.} & \textbf{Rec.} & \textbf{F1} \\
\midrule
\multicolumn{10}{l}{{Retrieval-based Methods }} \\
\midrule

NCE    & \textbf{90.30} & \textbf{78.90} & \textbf{84.22} & \underline{84.10} & 80.60 & \underline{82.31} & \multirow{5}{*}{} & &  \\
Text2TTP    & 51.59 & 21.36 & 30.22 & 74.76 & 74.65 & 74.70 & \\
BM25            & 67.86 & 64.74 & 66.26 & 32.54 & 32.48 & 32.51 & & \multicolumn{1}{c}{N/A} &  \\
RankGPT         & 61.93 & 58.56 & 60.20 & 59.40 & 59.33 & 59.37 &  \\
\textit{Our Re-Ranker}   & 64.69 & 61.43 & 63.02 & {85.46} & {85.29} & {85.37} & \\
\midrule
\multicolumn{10}{l}{{Generative Models}} \\
\midrule
GPT-4o          & 38.28 & 49.98 & 43.35 & 51.42 & 64.04 & 57.04 & 20.91 & 32.96 & 25.59 \\
\ \ w/ CoT+Ref      & 43.52 & 67.20 & 52.83 & 51.84 & 78.47 & 62.43 & 38.19 & 48.67 & 42.80 \\
DeepSeek v3    & 43.74 & 65.69 & 52.51 & 50.87 & 78.13 & 61.62 & 40.17 & 46.89 & 43.27 \\
\ \ w/ CoT+Ref  & 43.68 & 66.36 & 52.68 & 51.55 & 75.86 & 61.39 & 36.39 & 49.25 & 41.85 \\
Ministral 8B    & 7.68  & 31.71 & 12.36 & 7.07  & 30.79 & 11.50 & 8.47  & 19.63 & 11.84 \\
\ \ w/ CoT+Ref  & 14.94 & 26.21 & 19.03 & 16.58 & 29.29 & 21.17 & 16.88 & 17.17 & 17.02 \\
IntelEx             & 60.67 & 70.71 & 65.31 & 61.13 & 75.07 & 67.39 & 48.03 & 41.88 & 44.74 \\
\midrule
\multicolumn{10}{l}{{RAG Models}} \\
\midrule
GPT-4 (RAG)         & 55.50 & 70.64 & 62.16 & 71.34 & 88.06 & 78.82 & 47.49 & \underline{55.76} & \textbf{51.30} \\
DeepSeek v3 (RAG)   & 54.59 & \underline{77.36} & 64.01 & 66.43 & \underline{91.57} & 77.00 & 40.94 & \textbf{60.86} & 48.95 \\
Ministral 8B (RAG)  & 51.88 & 57.61 & 54.60 & 61.40 & 69.81 & 65.34 & 43.21 & 35.23 & 38.81 \\
{\bf \model} & \underline{76.00} & 72.14 & \underline{74.02} & \textbf{91.11} & \textbf{91.06} & \textbf{91.09} & \textbf{75.16} & {37.67} & \underline{50.19} \\
\bottomrule
\end{tabular}
}
\vspace{-8pt}
\caption{Results in \emph{technique} prediction. CoT+Ref: Chain-of-Thought w/ Reflection. The num of predicted labels are fixed for ranking models while generative models determines at runtime, hence compared in  Section \ref{sec:single-vs-multi-label-comparision}. 
}
\label{tab:combined-technique-level-enhanced}
\vspace{-10pt}
\end{table*}

\begin{table*}[t]
\centering
\small
\resizebox{0.85\linewidth}{!}{%
\begin{tabular}{l ccc | ccc | ccc}
\toprule
\multirow{2}{*}{\textbf{Model}} & \multicolumn{3}{c|}{\textbf{Tram (Single-label)}} & \multicolumn{3}{c|}{\textbf{Procedures (Single-label)}} & \multicolumn{3}{c}{\textbf{Expert (Multi-label)}} \\
\cmidrule(lr){2-10}
 & \textbf{Prec.} & \textbf{Rec.} & \textbf{F1} & \textbf{Prec.} & \textbf{Rec.} & \textbf{F1} & \textbf{Prec.} & \textbf{Rec.} & \textbf{F1} \\
\midrule
\multicolumn{10}{l}{{Retrieval-based Models}} \\
\midrule
NCE             & \textbf{77.00} & 65.80 & \textbf{70.96} & \underline{75.70} & 71.88 & 73.74 & \multirow{5}{*}{} & &  \\
Text2TTP        & 42.62 & 40.41 & 41.49 & 71.08 & 70.94 & 71.01 &  \\
BM25            & 48.41 & 46.56 & 47.47 & 24.17 & 24.17 & 24.17 & & \multicolumn{1}{c}{N/A} &  \\
RankGPT         & 43.03 & 40.97 & 41.97 & 51.19 & 51.12 & 51.16 & \\
Our Re-Ranker   & 50.76 & 48.45 & 49.58 & 84.21 & 83.98 & 84.10 & \\
\midrule
\multicolumn{10}{l}{{Generative Models}} \\
\midrule
GPT4o          & 27.62 & 36.34 & 31.38 & 42.42 & 55.04 & 47.91 & 16.69 & 24.26 & 19.77 \\
\ \ w/ CoT+Ref       & 32.33 & 49.91 & 39.24 & 43.90 & 68.33 & 53.46 & 32.24 & 37.75 & 34.78 \\
DeepSeek v3      & 30.97 & 47.07 & 37.36 & 40.98 & 64.23 & 50.04 & 34.60 & 35.87 & 35.22 \\
\ \ w/ CoT+Ref  & 33.28 & 51.47 & 40.42 & 41.47 & 61.63 & 49.58 & 30.79 & 37.14 & 33.67 \\
Ministral 8B     & 3.72  & 21.99 & 6.37  & 3.25  & 17.32 & 5.47  & 6.65  & 14.72 & 9.17 \\
\ \ w/ CoT+Ref  & 10.96 & 21.34 & 14.48 & 7.17  & 13.53 & 9.37  & 12.74 & 11.60 & 12.14 \\

IntelEx              & 53.09 & 63.33 & 57.76 & 53.07 & 67.77 & 59.53 & 43.55 & 33.52 & 37.88 \\
\midrule

\multicolumn{10}{l}{{RAG Models}} \\
\midrule
GPT4o (RAG)         & 39.29 & 52.84 & 45.07 & 64.11 & 81.63 & \underline{71.82} & \underline{41.77} & \underline{45.87} & \textbf{43.73} \\
DeepSeek v3 (RAG)   & 39.31 & 58.54 & 47.04 & 59.72 & \underline{86.47} & 70.65 & 35.91 & \textbf{48.06} & 41.11 \\
Ministral 8B (RAG)  & 34.94 & 40.86 & 37.67 & 53.41 & 63.75 & 58.12 & 32.90 & 28.24 & 30.39 \\
{\bf \model} & \underline{72.69} & \textbf{68.74} & \underline{70.66} & \textbf{91.11} & \textbf{88.09} & \textbf{88.11 }& \textbf{70.06} & 30.21 & \underline{42.22} \\
\bottomrule
\end{tabular}
}
\vspace{-8pt}
\caption{Performance Comparison for \emph{sub-technique} prediction task (in percentage). Note: CoT+Ref: Chain-of-Thought with Reflection. Retrieval-based methods are not applicable for the multi-label Expert dataset. 
}
\label{tab:combined-subtech-level-enhanced}
\vspace{-10pt}
\end{table*}

\section{Experiment Setup}
\label{sec:exp_setup}

\subsection{Data and Implementation}
\label{subsec:datasets_implementation}

We assess the capability of \modelnospace\ to accurately map threat behaviors to \emph{(sub-)technique} IDs.
We consider both single-label (\(l = 1\)) and multi-label (\(l > 1\)) prediction setups. Following previous works, we evaluate on three publicly available benchmark datasets: Tram \cite{tram:2023} as a single-label dataset, and the Procedures and Expert datasets \cite{nguyen-etal-2024-noise-huawei} representing single-label and multi-label settings, respectively. We report performances on the test sets of these datasets, training our model on the combined training sets.
Rather than developing separate models for \emph{technique} and \emph{sub-technique} prediction, we train a single model for \emph{sub-technique} prediction. This is motivated by the fact that \emph{sub-technique} annotations provide a more fine-grained representation that inherently includes the broader \emph{technique} identifier (e.g., in T1050.001, T1059 is the \emph{technique} and 001 is the \emph{sub-technique}). When evaluating for \emph{technique} prediction, we simply truncate the \emph{sub-technique} component.
As the retriever \(R\), we use BM25 with \(K=40\) and \(k=3\). For the frozen re-ranker \(\mathcal{R}\), we employ DeepSeek v3 \cite{liu2024deepseek} (with temperature set to 0) , processing retrieved candidates in batches of 40 with an overlap of 20. The trainable generator \(\mathcal{G}\) is implemented using an 8B Ministral Instruct model \cite{Ministral8BInstruct2410}. Fine-tuning is performed with LLaMa-Factory using LoRA \cite{hu2021lora}, with a learning rate of \(10^{-4}\), LoRA rank \(r=8\), and \(\alpha=4\). For generation, we use a sampling temperature of 0.7, a top-\(p\) value of 0.1, and a context length of 2048 tokens. Our source code, datasets, and models are publicly available on GitHub\footnote{\url{https://github.com/qcri/TechniqueRAG}}.

\subsection{Baselines and Evaluation Metrics}
\label{subsec:metrics}

\noindent\textbf{Retrieval/Ranking-only Methods}\quad
These include state-of-the-art approaches that rely solely on retrieval and re-ranking w/o using generative models. We compare w/ NCE~\cite{nguyen-etal-2024-noise-huawei} for contrastive domain-specific learning, Text2TTP~\cite{kumarasinghe2024semantic}, which combines bi-encoder semantic search w/ cross-encoder re-ranking, underlying BM25 retriever baseline, and RankGPT~\cite{sun-etal-2023-chatgpt} re-ranking framework that uses same BM25 retrievals. As NCE is not released we report from ~\cite{nguyen-etal-2024-noise-huawei}.

\noindent\textbf{Generation-based Methods}
 \textit{Direct Generation:} We evaluate against powerful LLMs including GPT-4, DeepSeek V3, and Ministral 8B. For each model, we implement both direct prompting and chain-of-thought approaches with self-reflection~\cite{shinn2024reflexion}.
 \textit{Retrieval-Augmented Generation:} We compare against IntelEX~\cite{xu2024intelex}, a hybrid retrieval and LLM-judge approach. Additionally, we implement retrieval-augmented versions of the above LLMs using text and identical exemplars from our retrieved and re-ranked pairs ($\mathcal{C}_x$).

\noindent\textbf{Evaluation Metrics}\quad
Following previous works, we evaluate performance on two settings: (i) for single-label {\em technique} and {\em sub-technique} prediction task, we use standard \textbf{Precision}, \textbf{Recall}, and \textbf{F1} scores; (ii) for multi-label tasks, we adopt a differentiated evaluation protocol. Our evaluation consists of: (1) End-to-End Evaluation: comparing our model's adaptive label predictions with generative baselines, as both can dynamically determine the optimal number of labels—a capability that retriever-only methods lack; and (2) Ranking Analysis: evaluating our re-ranker against all retriever or ranking methods using standard ranking metrics (Precision, Recall, and F1) at k=\{1,3\}.

\section{Results and Analysis}
\label{sec:result-analysis}

\subsection{{\em Technique}-Level Performance}

Table \ref{tab:combined-technique-level-enhanced} reports the performance of various models on the technique prediction task across three datasets with increasing diversity: Tram (single-label with 198 unique), Procedures (single-label with 488 unique), and Expert (multi-label with 290 unique) {\em techniques} and {\em sub-techniques}. Among retrieval-based methods, NCE achieves the highest F1 score on Tram (84.22\%), reflecting its strength in a constrained label space. However, as the diversity increases, NCE’s performance drops markedly—for example, on Procedures it only reaches an F1 of 82.31\%

In contrast, our proposed \model excels consistently. On Procedures, \model attains an F1 score of 91.09\%, demonstrating its superior ability to generalize in a high-diversity, single-label setting. Although on the Expert dataset proprietary model GPT-4o (RAG) achieves a marginally higher F1 (51.30\% vs. \modelnospace’s 50.19\%), the overall performance averaged across the three datasets signifies the effectiveness of our open-source framework. When we compute the average F1 score, \model achieves approximately 80.76\%, compared to only about 58.11\% for GPT-4o (RAG). This substantial improvement underscores that our model is more robust, particularly in handling diverse and complex adversarial scenarios.

\subsection{{\em Sub-Technique}-Level Performance}

Table \ref{tab:combined-subtech-level-enhanced} presents the results at the {\em sub-technique} level. A similar trend is observed. At finer granularity, our method maintains dominance on Procedures (our F1 88.11 vs NCE's 73.74) while matching Tram's performance gap (F1 70.66 vs NCE's 70.96). This again posits our effectiveness for complex and robust threat annotation. 
The performance gap between our model and other generative and RAG baselines widens further at the {\em sub-technique} level. While GPT-4o (RAG) achieves a slightly higher F1 score on the Expert dataset (43.73 vs our 42.22), the overall results across all datasets demonstrate that our approach generalizes more effectively to complex, high-diversity environments.


\begin{table*}[t]

\centering
\small
\begin{tabular}{@{}l cc cc cc | cc cc cc@{}}  
\toprule
\multirow{2}{*}{\textbf{Model}} & 
\multicolumn{6}{c}{\textbf{Technique Level}} & \multicolumn{6}{c}{\textbf{Sub-Technique Level}} \\
\cmidrule(lr){2-7} \cmidrule(lr){8-13}
 & \multicolumn{3}{c}{\textbf{@1}} & \multicolumn{3}{c}{\textbf{@3}} & \multicolumn{3}{c}{\textbf{@1}} & \multicolumn{3}{c}{\textbf{@3}} \\
\cmidrule(lr){2-4} \cmidrule(lr){5-7} \cmidrule(lr){8-10} \cmidrule(lr){11-13}
 & P & R & F1 & P & R & F1 & P & R & F1 & P & R & F1 \\
\midrule
NCE          & \textbf{74.5} & 23.6 & \underline{35.9} & — & — & \underline{48.3} & \textbf{73.1} & 18.2 & 29.1 & — & — & \underline{39.9} \\
Text2TTP     & 53.5 & \underline{26.1} & 35.1 & \underline{37.4} & \underline{49.1} & 42.4 & 49.0 & \underline{21.3} & \underline{30.2} & 34.4 & \underline{39.7} & 36.8 \\
BM25         & 51.6 & 21.4 & 30.2 & 35.5 & 40.4 & 37.8 & 45.9 & 15.6 & 23.3 & 31.0 & 29.9 & 30.5 \\
RankGPT      & 56.7 & 25.3 & 34.9 & \underline{37.4} & 46.6 & 41.5 & 49.7 & 19.8 & 28.4 & \underline{34.8} & 37.8 & 36.3 \\
Our Re-Ranker & \underline{71.3} & \textbf{35.3} & \textbf{47.2} & \textbf{44.6} & \textbf{59.9} & \textbf{51.1} & \underline{66.9} & \textbf{29.0} & \textbf{40.5} & \textbf{47.1} & \textbf{54.2} & \textbf{50.4} \\
\bottomrule
\end{tabular}
\vspace{-8pt}
\caption{Performance of Ranking Methods on Expert Dataset (Multi-Label). '-' refers to results not reported. 
}
\label{table:rerank-combined}
\vspace{-10pt}
\end{table*}

\subsection{Single-Label versus Multi-Label Settings}
\label{sec:single-vs-multi-label-comparision}

Our experiments in Table \ref{tab:combined-technique-level-enhanced} and \ref{tab:combined-subtech-level-enhanced} reveal that 
multi-label prediction poses significant challenges in comparison to single-label. For example GPT-4o achieves an F1 score of 47.91 in Procedure while only 19.77 in Expert).  While retrieval augmented generation enhances all generative models, gains in open-source  LLMs remain low such as using Ministral RAG without our finetuning scores an F1 of 30.39 in Expert. Adapting to the domain \model boosts it up to 42.22 tailing the RAG score of GPT-4o's 43.73. Furthermore, in Table \ref{table:rerank-combined} we compare all the ranking based models with our re-ranker framework--showing a large margin gains over all. These significant F1 improvements both in single and multi-label setup confirm the effectiveness of our model in real-world scenarios. 


\subsection{Ablation Study}
\noindent{\bf Enhancement with Our Re-Ranker}
Our comprehensive evaluation in Table \ref{tab:combined-technique-level-enhanced}, \ref{tab:combined-subtech-level-enhanced} and \ref{table:rerank-combined} clearly indicate that our re-ranker not only outperforms all the ranking based methods but also enhance the overall end-to-end performances. In addition to our model, all the generative models (e.g., DeepSeek) in RAG setups using our re-ranked exemplars achieves notable gains over their direct or CoT+Ref inferences. 

\begin{figure}[h]
    \centering
    \begin{subfigure}[b]{0.48\textwidth}
        \centering
        \includegraphics[width=\textwidth]{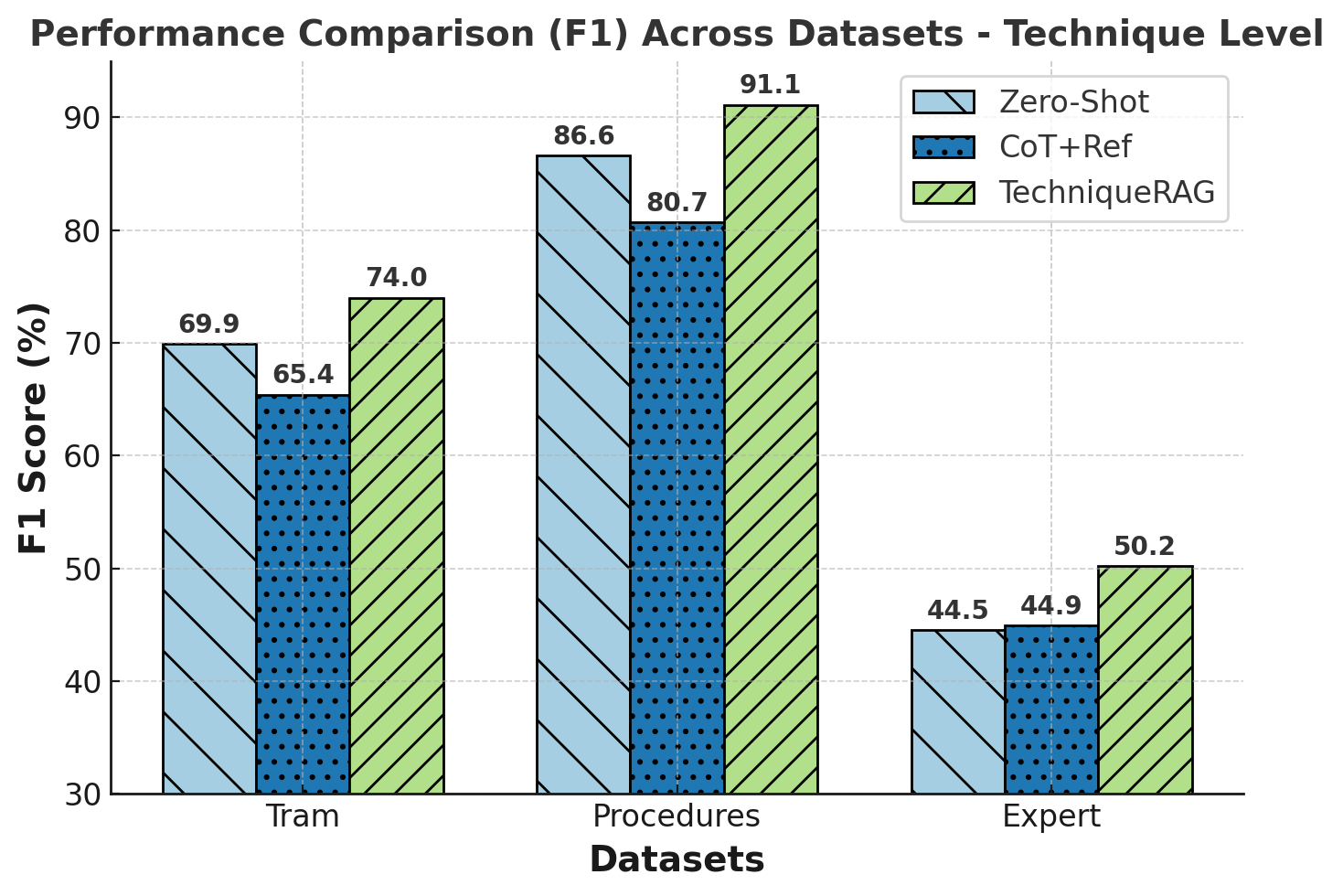}  
        \caption{Performance Comparison (F1) - Technique Level}
        \label{fig:technique_level}
    \end{subfigure}
    \hfill
    \begin{subfigure}[b]{0.48\textwidth}
        \centering
        \includegraphics[width=\textwidth]{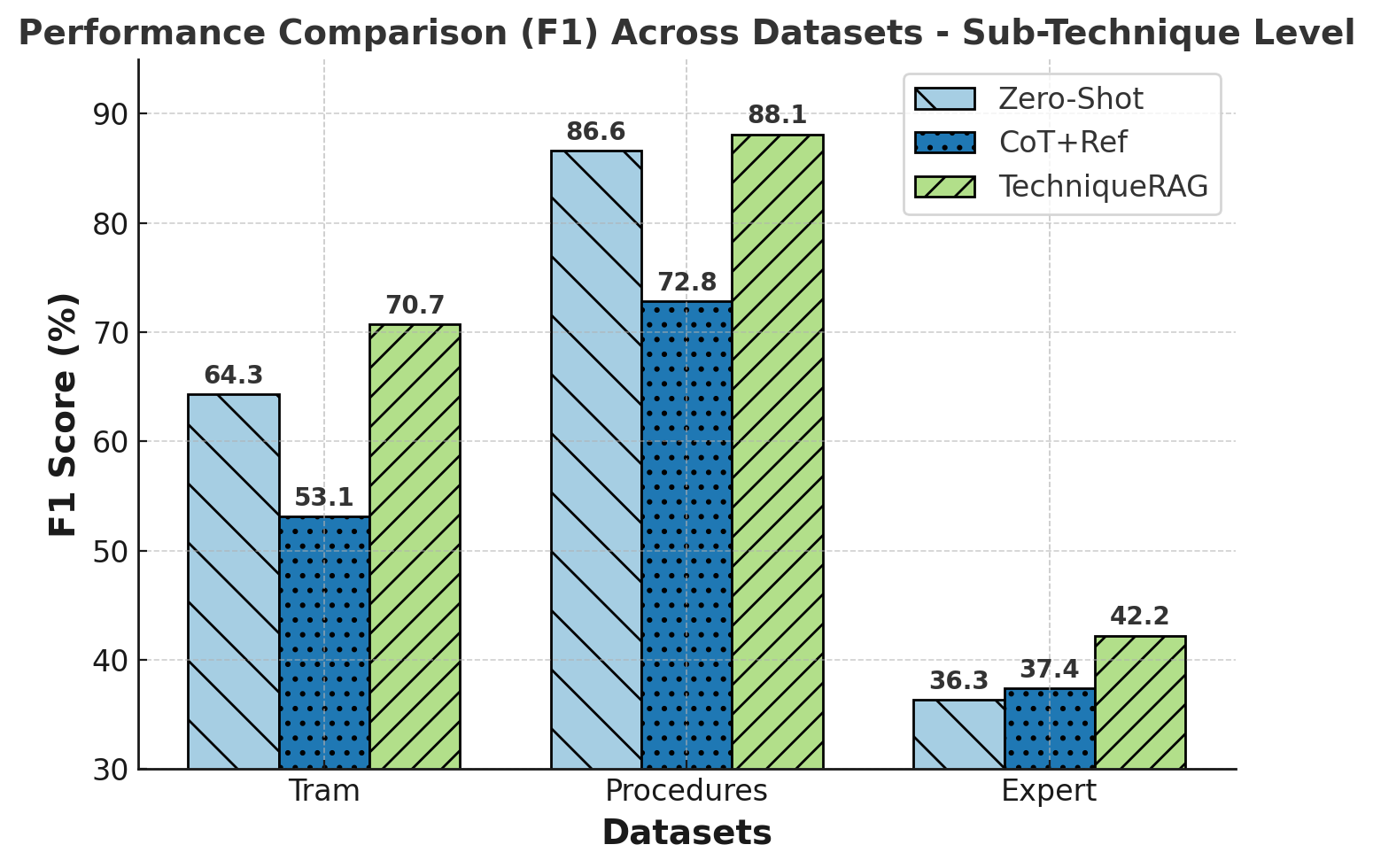}  
        \caption{Performance Comparison (F1) - Sub-Technique Level}
        \label{fig:sub_technique_level}
    \end{subfigure}

    \caption{ F1 scores  for different fine-tuning methods.}
    \label{fig:sft-comparison}
    \vspace{-10pt}
\end{figure}

\noindent{\bf Gains over Other Fine-Tuning Methods}
We also validate the effectiveness of our RAG-based domain adaptation methods over zero-shot and CoT+Ref based methods. For zero-shot, we fine-tune our same Ministral model on the same training data but without exemplars and for CoT+Ref based methods we followed the Alpaca approach \cite{alpaca}: using the same training data, we prompt the DeepSeek v3 model with CoT+Ref instructions to synthesize new training examples, which we then use to fine-tune the Ministral model. Results in Fig \ref{fig:sft-comparison} shows our approach achieves the highest gain in both target tasks in all benchmark datasets.

\subsection{Qualitative Analysis}
\label{subsec:error_analysis}
\noindent\textbf{Running Example.} We provide in Appendix \ref{apdx:running_example} a concrete example from Expert dataset that shows the predictions of our Re-Ranker and how \modelnospace~ generator improved it. We also provide examples of the prompts in RankGPT and ours with detailed responses with our re-ranker LLM DeepSeek V3 in Appendix \ref{apdx:prompts}. 

\noindent\textbf{Error Analysis}
Analysis reveals few challenges:

\textit{Under-prediction.} The model often captures primary techniques while missing related techniques in the same attack pattern (e.g., identifying T1055 but missing associated techniques like T1106)

\textit{Contextual errors.}
(i) Confusion between similar techniques within the same tactic family specially for \emph{Command and Scripting Interpreter} techniques (T1059.*)
(ii) Missing implicit or contextual techniques not explicitly stated
(iii) Difficulty capturing logical relationships between techniques

\textit{Hierarchical issues.} Struggles with parent-child technique relationships and sometimes generates invalid sub-technique IDs.

\textit{Re-ranker limitations.} Missed {\em techniques} due to ambiguous queries and compound statements, affecting the Generator through propagation.

\textit{Technique similarity.} Challenges in distinguishing between {\em techniques} with overlapping descriptions and keywords (e.g., phishing-related techniques T1598.003, T1566.002, T1204.001).

\textit{Class Imbalance Effects.} The severe data imbalance fundamentally impacts model performance - only 47 out of 203 {\em techniques} (23.2\%) have more than 50 training samples. Techniques with abundant data show high precision and recall, while rare {\em techniques} suffer from both misclassification and under-prediction.

We present detailed analysis in Appendix \ref{sec:appendix:error-analysis}.

\section{Conclusion}
\label{sec:conlcusion}

Annotating threat intelligence texts with adversarial {\em techniques} from the MITRE ATT\&CK framework is a manual and time-intensive task that security analysts must perform daily. Its automation requires methods capable of accurately identifying {\em techniques} and {\em sub-techniques} across hundreds of possibilities while handling complex security terminology, diverse text formats, and limited labeled data.
We introduce \modelnospace\, a retrieval-augmented fine-tuning approach designed to tackle these challenges effectively. Our comparative analysis demonstrates that \modelnospace\ establishes a new state-of-the-art, outperforming both semantic ranking models and other LLM-based methods in adversarial technique annotation.

\section{Limitations}
Obtaining large, balanced parallel datasets of threat descriptions and ground truth technique annotations remains a significant challenge due to the reliance on domain expertise for accurate annotation. Although our approach mitigates data scarcity, two key limitations may impact performance:
\begin{enumerate}
    \item \textbf{Limited Technique Coverage:} Coverage of techniques is often insufficient. Even the MITRE ATT\&CK knowledge base lacks procedural examples for many techniques and sub-techniques.

    \item \textbf{Sparse Technique Annotations:} Existing datasets typically contain very few technique annotations per example, with many instances in our data having only a single technique label. During fine-tuning, this bias toward minimal technique labeling limited our method’s ability to generalize effectively. To mitigate this, we oversampled examples with multi-label technique annotations. However, our method rarely assigned more than two technique labels per input query, leading to low recall, particularly on the Expert dataset, which consists almost exclusively of multi-label examples.

    \item \textbf{Annotation Inconsistencies} Some model predictions marked as errors represent valid technical interpretations not included in gold standard annotations. For example, the following sentence: \textit{``SMOKEDHAM was observed using UltraVNC to establish a connection to the IP address and port pair 81.91.177[.]54[:]7234 that has been observed in past UNC2465 intrusions.''} had \textit{T1571: Non-Standard Port} as the only ground truth label. However, if we analyze it carefully, we see that the threat actor used UltraVNC, so \textit{T1021.005: Remote Services - VNC} exists in the given description. Our model correctly predicted it, but missed the \textit{T1571: Non-Standard Port}. This highlights challenges in maintaining consistent annotation standards for complex attack patterns.
\end{enumerate}

\section*{Acknowledgments}

The authors declare the use of AI Assistants in editing the text, fixing the typos, and generating the prompts and figures.

\section*{Appendix}
\appendix

\section{Running Example}
\label{apdx:running_example}
In Figure \ref{fig:running_example}, the given text describes various execution techniques used by attackers, such as launching executables and DLLs in memory, leveraging schtasks.exe to modify task schedules, and executing PowerShell commands. The ground truth labels identify the relevant MITRE ATT\&CK techniques: T1059.001 (PowerShell) is explicitly mentioned in ``\textit{The ability to launch custom PowerShell commands,}'' T1053 (Scheduled Task/Job) and its sub-technique T1053.005 (Scheduled Task) are referenced in ``\textit{The ability to leverage schtasks.exe to add or modify task schedules.}'' The RAG output provided partial alignment with the ground truth labels but included T1218.011 (Rundll32), which is not explicitly mentioned. However, this retrieval step helped the LLM Output by reinforcing correct predictions ensuring T1059.001 and T1053.005 were present while also introducing T1053. Additionally, the LLM expanded the predictions to include T1071.001 (Web Protocols), which does not appear relevant in this case. 

\begin{figure}[!hbp]
\begin{tcolorbox}[title=Example of \modelnospace~ Output, colback=blue!5!white, colframe=blue!25!black, fonttitle=\footnotesize, boxrule=0.2mm, sharp corners]
\small
The ability to launch a .exe or .dll file in memory. The ability to leverage \hlred{``schtasks.exe''} to \hlpurple{add or modify task schedules}. The ability to launch \hlorange{custom PowerShell commands}. The ability to leverage a \hlteal{standalone executable to load the DLL} if the attacker otherwise has no way of doing so.\\
---------------------------------------------------------------  \\
\textbf{Ground Truth Labels:} \\
1. \hlorange{T1059.001: Command and Scripting Interpreter: PowerShell} \\
2. \hlred{T1053: Scheduled Task/Job} \\
3. \hlpurple{T1053.005: Scheduled Task/Job: Scheduled Task} \\
---------------------------------------------------------------  \\
\textbf{RAG Output (Our Re-ranker):} \\
1. \textcolor{gray}{T1218.011: System Binary Proxy Execution: Rundll32} \\
2. \hlorange{T1059.001: Command and Scripting Interpreter: PowerShell} \\
3. \hlpurple{T1053.005: Scheduled Task/Job: Scheduled Task} \\
---------------------------------------------------------------  \\
\textbf{LLM Output (\modelnospace):} \\
1. \hlorange{T1059.001: Command and Scripting Interpreter: PowerShell} \\
2. \hlred{T1053: Scheduled Task/Job} \\
3. \hlpurple{T1053.005: Scheduled Task/Job: Scheduled Task} \\
4. \textcolor{gray}{T1071.001: Application Layer Protocol: Web Protocols} \\
\end{tcolorbox}
\vspace{-12pt}
\caption{Example of MITRE ATT\&CK {\em techniques} and {\em sub-techniques} highlighted in text with corresponding colored (implicit) indicators. IDs with "." denote sub-{\em techniques} (e.g., T1059.001). Greyed-out IDs indicate incorrect predictions.}
\label{fig:running_example}
\vspace{-20pt}
\end{figure}

\section{Data Statistics}
\label{sec:appendix:data-stat}

Tables \ref{tab:dataset} and \ref{tab:datasets_st} shows the details of the employed datasets.
The Expert split consists of actual sentences from full reports published by threat intelligence vendors. These sentences are multi-label, meaning they can be associated with multiple MITRE  ATT\&CK techniques. In contrast, the Tram split contains incomplete sentences, such as ``\textit{opens cmd.exe on the victim}'', ``\textit{searches for specified files}'', or ``\textit{icacls . /grant Everyone:F /T /C /Q}'', often presenting isolated technique references without sufficient context. Tram is single-label, meaning each sentence corresponds to only one technique. The Procedures split, extracted from the MITRE knowledge base, consists of complete sentences that summarize a single technique mentioned in a report. These sentences provide structured descriptions of attack techniques but are also single-label. In total, the training splits contain 499 unique techniques, covering approximately 78\% of the 637 techniques available in the MITRE ATT\&CK Enterprise Framework.


\begin{table}[!htbp]
\caption{Dataset Statistics}
\label{tab:dataset}
\scalebox{0.9}{
\resizebox{\columnwidth}{!}{%
\begin{tabular}{lccc}
\toprule
\textbf{Dataset} & \textbf{Split} & \textbf{Avg Word Count} & \textbf{Data} \\ 
\midrule
\multirow{2}{*}{\textbf{Expert}}             & Train & 38.00 & 472   \\  
                                             & Test  & 71.42 & 158  \\ 
\midrule
\multirow{2}{*}{\textbf{Procedures}}         & Train & 13.36 & 10,999 \\ 
                                             & Test  & 13.43 & 1768 \\  
\midrule
\multirow{2}{*}{\textbf{Tram}}               & Train & 2.94 & 3469  \\  
                                             & Test  & 21.22 & 726 \\  
\bottomrule
\end{tabular}
}
}
\end{table}

\begin{table}[!htbp]
\caption{Dataset statistics. \emph{S+T} denotes the joint count of techniques and sub-techniques.}
\scalebox{0.9}{
\resizebox{\columnwidth}{!}{%
\begin{tabular}{lrcccc}
\toprule
\centering 

\multirow{2}*{\textbf{Dataset}} & 
\multirow{2}*{\textbf{Texts}} & 
\multirow{2}*{\textbf{S+T}} & 
\textbf{Tech-} & 
\textbf{Avg. \#} & 
\textbf{Avg. \#} \\

&
&
&
\textbf{niques} &
\textbf{Labels} &
\textbf{Tokens} \\

\midrule
\textit{TRAM}               & 4797       & 193   & 132    & 1.16       & 23         \\
\textit{Procedures}         & 11723      & 488   & 180    & 1.00       & 12         \\
\textit{Expert}             & 695        & 290   & 151    & \textbf{1.84 }      & \textbf{72}  \\
\bottomrule
\end{tabular}
}
}
\label{tab:datasets_st}
\end{table}

\section{Error Analysis}
\label{sec:appendix:error-analysis}

\noindent \textbf{Common Errors.} Analysis of the prediction errors reveals several systematic patterns in MITRE ATT\&CK technique classification. The most frequent error type involves under-prediction, where the model identifies only the most prominent technique while missing other techniques that are part of the same attack pattern. For example, when analyzing process injection scenarios, the model often identifies the primary technique (\emph{T1055: Process Injection}) but fails to capture associated techniques like \emph{T1106: Native API} or specific sub-techniques like \emph{T1055.001: Dynamic-link Library Injection}. Another common pattern involves confusing similar techniques within the same tactic family, particularly between various \emph{Command and Scripting Interpreter} techniques (T1059.*). The model also demonstrates a tendency to miss contextual techniques that are implied but not explicitly stated in the text, such as failing to identify \emph{T1082: System Information Discovery} when enumeration of system resources is described as part of a larger operation. Additionally, there is a notable pattern of missing data staging and encoding techniques (T1074, T1132) when they are described as part of exfiltration workflows.

\subsection {Contextual Inference Failures.}
The model demonstrates limitations in capturing implicit relationships, often missing techniques that are logical precursors or consequences of explicitly described actions. It frequently identifies primary techniques while missing related concurrent techniques within the same attack pattern.

\textbf{Class Imbalance Effects.} The severe data imbalance fundamentally impacts model performance - only 47 out of 203 techniques (23.2\%) have more than 50 training samples. Techniques with abundant data show high precision and recall, while rare techniques suffer from both misclassification and under-prediction.
    

\subsection{Dependency on Re-ranker}
In several cases, some \emph{(sub-)}techniques are omitted, likely due to ambiguous language in the query or an overemphasis on the most actionable part of a compound query for example. This, error further propagate to our Generator, which uses output from Re-ranker as few-shot examples.
    
\subsection{Similar Techniques}
Several techniques within the MITRE ATT\&CK framework share significant similarities and often use overlapping keywords, which can influence our initial BM25 rankings. For instance, T1598.003, T1566.002, and T1204.001 are all phishing-related techniques that have similar descriptions with minor distinctions.

\section{Different Domain Adaptation Methods}
\label{sec:appendix:domain-adaptation-methods}

\onecolumn

\begin{table*}[t]
\caption{Performance Comparison of \model Across Domain Adaptation techniques Levels (Percentage Scores)}
\label{tab:combined-granularity}
\centering
\small
\resizebox{\linewidth}{!}{%
\begin{tabular}{@{}lcccccc|cccccc|cccccc@{}}
\toprule
\multirow{2}{*}{\textbf{Model}} & 
\multicolumn{6}{c}{\textbf{Tram}} & \multicolumn{6}{c}{\textbf{Procedures}} & \multicolumn{6}{c}{\textbf{Expert}} \\
\cmidrule(lr){2-7} \cmidrule(lr){8-13} \cmidrule(lr){14-19}
 & \multicolumn{3}{c}{Technique} & \multicolumn{3}{c}{Sub-Technique} & \multicolumn{3}{c}{Technique} & \multicolumn{3}{c}{Sub-Technique} & \multicolumn{3}{c}{Technique} & \multicolumn{3}{c}{Sub-Technique} \\
\cmidrule(lr){2-4} \cmidrule(lr){5-7} \cmidrule(lr){8-10} \cmidrule(lr){11-13} \cmidrule(lr){14-16} \cmidrule(lr){17-19}
 & P & R & F1 & P & R & F1 & P & R & F1 & P & R & F1 & P & R & F1 & P & R & F1 \\
\midrule
Zero-shot & 
65.7 & \textbf{74.9} & 69.9 & 60.0 & \textbf{69.2} & 64.3 & 85.8 & 87.4 & 86.6 & 85.8 & 87.4 & 86.6 & 41.3 & 48.3 & 44.5 & 35.1 & \textbf{37.6} & 36.3 \\

+ CoT+Ref & 
64.1 & 66.7 & 65.4 & 52.0 & 54.3 & 53.1 & 77.4 & 84.3 & 80.7 & 69.1 & 76.9 & 72.8 & 48.6 & 41.7 & 44.9 & 42.4 & 33.5 & 37.4 \\

\model & 
\textbf{76.0} & 72.1 & \textbf{74.0} & \textbf{72.7} & 68.7 & \textbf{70.7} & \textbf{91.1} & \textbf{91.1} & \textbf{91.1} & \textbf{91.1} & \textbf{88.09} & \textbf{88.11} & \textbf{75.2} & 37.7 & \textbf{50.2} & \textbf{70.1} & 30.2 & \textbf{42.2}\\
\bottomrule
\end{tabular}
}
\vspace{0.5em}
\end{table*}

\section{Prompts}
\label{apdx:prompts}

\begin{figure*}[h]
    \centering
    \begin{tcolorbox}[
        title=Vanilla RankGPT Prompt and Output,
        colback=blue!5!white,
        colframe=blue!25!black,
        fonttitle=\footnotesize,
        boxrule=0.2mm,
        sharp corners,
        width=0.95\columnwidth
    ]
    \label{fig:vanilla_rankgpt_prompt}
    \small
    \textbf{\# System Prompt:} \\
    You are RankGPT, an intelligent assistant that can rank passages based on their relevancy to the query. \\\\
    \textbf{\#\# Objectives:} \\
    I will provide you with \{num\} passages, each indicated by a number identifier [ ]. \\
    Rank the passages based on their relevance to the query. \\\\
    \textbf{\#\# Given Passages:} \\
    \{Passage 1: Description\} \\
    \{Passage 2: Description\} \\
    $\vdots$ \\
    \{Passage n: Description\} \\\\
    \textbf{\#\# Query:} \\
    Monero miner scripts are downloaded from TeamTNT’s server and piped to \texttt{bash} using an SSH session on the underlying host as the \texttt{root} user by supplying the private key from \texttt{/tmp/TeamTNT}. Later, the private key \texttt{/tmp/TeamTNT} is removed as well. \\\\
    \rule{\columnwidth}{0.1mm}\\\\
    \textbf{\#\# Output} \\
    
\textcolor{darkgreen}{T1552.004} $>$ \textcolor{darkgreen}{T1098.004} $>$ T1563.001 $>$ T1021 $>$ T1555.002 $>$ T1573.002 $>$ T1546.004 $>$ T1496 $>$ \textcolor{darkgreen}{T1059.004} $>$ T1611

    \end{tcolorbox}
    \caption{Example RankGPT prompt and its corresponding output. Green colored IDs are the correct ones.}
\end{figure*}

\begin{figure*}[t]
    \centering
    \begin{tcolorbox}[
        title=Re-ranker Prompt,
        colback=blue!5!white,
        colframe=blue!25!black,
        fonttitle=\footnotesize,
        boxrule=0.2mm,
        sharp corners,
    ]
    \label{app:reranker-prompt}
    \small
    \textbf{\# System Prompt} \\
    Act as an expert security analyst specializing in ranking a given list of MITRE ATT\&CK techniques by their relevance to a security query. \\\\
    \textbf{\#\# Objectives:} \\
    1. Determine if the given query describes an adversarial or attack behavior. \\
    2. If it does, identify and rank the most relevant techniques and sub-techniques from the provided list. \\\\
    \textbf{\#\# Instructions for Ranking:} \\
    \textbf{1. Break Down the Query:} \\
    - Decompose the security query into distinct attack steps. \\
    - Identify explicit and implied adversarial behaviors. \\
    \textbf{2. Match Techniques:} \\
    - Map each behavior to relevant techniques/sub-techniques. \\
    - Consider multiple relevant techniques. \\
    \textbf{3. Provide Explanation:} \\
    - Explain the connection between the query and the matched techniques. \\\\
    \textbf{\#\# Final Output Format:} \\
    After reasoning, output the final ranking as:
    [Technique A] > [Technique B] > [Technique C] > ... \\
    No variations or extra formatting allowed. \\\\
    \textbf{\#\# Given Techniques:} \\
    \{Technique 1: Description\} \\
    \{Technique 2: Description\} \\
    $\vdots$ \\
    \{Technique n: Description\} \\\\
    \textbf{\#\# Query:} \\
    Monero miner scripts are downloaded from TeamTNT’s server and piped to \texttt{bash} using an SSH session on the underlying host as the \texttt{root} user by supplying the private key from \texttt{/tmp/TeamTNT}. Later, the private key \texttt{/tmp/TeamTNT} is removed. \\\\
    \rule{\columnwidth}{0.1mm}\\\\
    \textbf{\#\# LLM Output} \\
    
The query describes an attack involving the following key steps and techniques:

Initial Access: The attacker gains access to the host via SSH using a private key stored in /tmp/TeamTNT. This aligns with T1552.004 (Unsecured Credentials: Private Keys) and T1021.004 (Remote Services: SSH).

Execution: The attacker downloads Monero miner scripts and pipes them to bash. This involves T1059.004 (Command and Scripting Interpreter: Unix Shell).

Persistence: The attacker uses SSH with a private key, which could imply persistence via T1098.004 (Account Manipulation: SSH Authorized Keys).

Resource Hijacking: The Monero miner script indicates T1588.001 (Obtain Capabilities: Malware) for downloading and executing the miner.

Key Techniques Identified:
T1552.004: Exploitation of private keys for SSH access.

T1021.004: Use of SSH for remote access.

T1059.004: Execution of scripts via Unix shell (bash).

T1098.004: Potential persistence via SSH authorized keys.

T1588.001: Downloading and executing Monero miner scripts (malware).

\textbf{Ranking Explanation}: \\
Passages [1], [2], [6], [7], [8], [40] are highly relevant as they directly address the techniques used in the query. Passages [3], [5], [9], [10], [11], [12], [13], [14], [15], [16], [17], [18], [19], [20], [21], [22], [23], [24], [25], [26], [27], [28], [29], [30], [31], [32], [33], [34], [35], [36], [37], [38], [39] are less relevant as they describe techniques not directly involved in the query.

    \textbf{Final Ranking:} \\
\textcolor{darkgreen}{T1552.004} $>$ \textcolor{darkgreen}{T1098.004} $>$ \textcolor{darkgreen}{T1021.004} $>$ \textcolor{darkgreen}{T1059.004} $>$ T1588.001 $>$ T1496 $>$ T1563.001 $>$ T1546.004 $>$ T1611 $>$ T1140

    \end{tcolorbox}
    \caption{Example re-ranker prompt and its corresponding output. Green colored IDs are the correct ones.}
\end{figure*}

\begin{figure*}[t]
    \centering
    \begin{tcolorbox}[
        title=Self-Relflection Prompt,
        colback=blue!5!white,
        colframe=blue!25!black,
        fonttitle=\footnotesize,
        boxrule=0.2mm,
        sharp corners,
        width=0.95\columnwidth
    ]
    \label{fig:reflection_prompt}
    \small
    Your task is to analyze a given text describing malware behavior, extract the associated MITRE ATT\&CK techniques to this text, explain their relevance to it. \\
    
    \textbf{\#\# Context:} \\
    The MITRE ATT\&CK framework is a globally-accessible knowledge base of adversary tactics and techniques based on real-world observations. It's used by cybersecurity professionals to better understand and defend against cyber threats. \\
    
    \textbf{\#\# Your Task:} \\
    Analyze the following text. Extract all the associated MITRE ATT\&CK technique and provide a detailed explanation of why each technique is relevant to the text. \\
    
    \textbf{\#\# Instructions:} \\
    Follow these steps to provide your analysis:\\
    1. <thinking>: Explain your thought process as you analyze the given cyber threat description. Identify key actions, tools, or methods mentioned that could correspond to MITRE ATT\&CK techniques in 30 words.\\
    2. <reflection>: Reflect on your initial analysis. Consider if you've missed any potential techniques or if any of your initial thoughts need revision. Think about the confidence level of your associations in 30 words.\\
    3. <output>: Based on your thinking and reflection, output the final list of MITRE ATT\&CK techniques as technique IDs and their names. For example:\\
    <output>\\
    - T1221: Template Injection\\
    - T1205.001: Traffic Signaling - Port Knocking\\
    </output>\\
    
    Ensure you use these exact tags (<thinking>, <reflection>, and <output>) in your response. \\
    
    \textbf{\#\# Output Format:} \\
    <thinking> \\
    Based on the given cyber threat description, I can identify several key actions and tools that correspond to MITRE ATT\&CK techniques: \\
    1. [Insert relevant observations from the text] \\
    2. [Continue with more observations] \\
    
    These observations suggest the following potential MITRE ATT\&CK techniques: \\
    - [List potential techniques with brief explanations] \\
    </thinking> \\
    
    <reflection>\\
    Upon reflection, I should consider the following:\\
    1. Are there any subtle indicators in the text that I might have overlooked: [Your answer in 20 words or less for question 1] \\
    2. Have I considered the full context of the attack, including potential preliminary or subsequent steps not explicitly mentioned? [Your answer in 20 words or less for question 2] \\
    3. Are there any techniques I've identified that might not be fully supported by the given information? [Your answer in 20 words or less for question 3] [Add any additional reflections or revisions to the initial analysis] 

    Confidence level: [State the confidence level in the identified techniques]  \\ 
    </reflection>  \\
    
    <output> 
    [List the final list of the extracted MITRE ATT\&CK techniques as technique IDs and their names.] \\
    </output>\\
    \end{tcolorbox}
    \caption{The employed prompt in self-reflection.}
\end{figure*}



\end{document}